\documentclass[twocolumn]{aastex62}
\usepackage{graphics,epsf}
\usepackage{amsmath}                
\usepackage{amsfonts}               
\usepackage{amssymb}                
\usepackage{epsfig}                 
\usepackage{graphicx}
\usepackage{float}
\usepackage{color}
\usepackage[para,online,flushleft]{threeparttable}

\newcommand{\cm}{{~\rm cm}}
\newcommand{\km}{{~\rm km}}
\newcommand{\s}{{~\rm s}}

\newcommand{\g}{{~\rm g}}

\newcommand{\erg}{{~\rm erg}}
\newcommand{\yr}{{~\rm yr}}

\begin{document}

\title{Inclined jets inside a common envelope of a triple stellar system}


\author{Ron Schreier}
\affiliation{Department of Physics, Technion, Haifa, 3200003, Israel;  ronsr@physics.technion.ac.il, shlomihi@tx.technion.ac.il, soker@physics.technion.ac.il}

\author{Shlomi Hillel}
\affiliation{Department of Physics, Technion, Haifa, 3200003, Israel;  ronsr@physics.technion.ac.il, shlomihi@tx.technion.ac.il, soker@physics.technion.ac.il}

\author[0000-0003-0375-8987]{Noam Soker}
\affiliation{Department of Physics, Technion, Haifa, 3200003, Israel;  ronsr@physics.technion.ac.il, shlomihi@tx.technion.ac.il, soker@physics.technion.ac.il}
\affiliation{Guangdong Technion Israel Institute of Technology, Shantou 515069, Guangdong Province, China}

\begin{abstract}
We conduct a three-dimensional hydrodynamical simulation to study the interaction of two opposite inclined jets inside the envelope of a giant star, and find that the jets induce many vortexes inside the envelope and that they efficiently remove mass from the envelope and form a very clumpy outflow. We assume that this very rare type of interaction occurs when a tight binary system enters the envelope of a giant star, and that the orbital plane of the tight binary system and that of the triple stellar system are inclined to each other. We further assume that one of the stars of the tight binary system accretes mass and launches two opposite jets and that the jets' axis is inclined to the angular momentum axis of the triple stellar system.
The many vortexes that the jets induce along the orbit of the tight binary system inside the giant envelope might play an important role in the common envelope evolution (CEE) by distributing energy in the envelope. The density fluctuations that accompany the vortexes lead to an outflow with many clumps that might facilitate the formation of dust.
This outflow lacks any clear symmetry, and it might account for very rare types of `messy' planetary nebulae and `messy' nebulae around massive stars.
On a broader scope, our study adds to the notion that jets can play important roles in the CEE, and that they can form a rich variety of shapes of nebulae around evolved stars. 
\end{abstract}


\section{INTRODUCTION}
\label{sec:intro}

Some early studies suggested that stellar binary interaction shapes the outflow from only a fraction of asymptotic giant branch (AGB) stars to form non-spherical planetary nebulae (PNe; e.g., \citealt{LivioShaviv1975, Bondetal1978, FabianHansen1979, Morris1981, Paczynski1985,IbenTutukov1989, BondLivio1990, Hanetal1995, Bond2000}). Hundreds of more recent studies, on the other hand, suggest that the majority, if not even all, PNe are shaped by binary interaction
(e.g., \citealt{Jonesetal2016, Chiotellisetal2016, Akrasetal2016, GarciaRojasetal2016, Jones2016, Hillwigetal2016a, Bondetal2016, Chenetal2016, Madappattetal2016, Alietal2016, Hillwigetal2016b, Jonesetal2017, JonesBoffin2017b, Barker2018, BondCiardullo2018, Bujarrabaletal2018, Chenetal2018, Danehkaretal2018, Franketal2018, GarciaSeguraetal2018, Hillwig2018,  MacLeodetal2018, Miszalskietal2018PASA, Sahai2018ASPC, Wessonetal2018, Desmursetal2019, Jones2019H, Kimetal2019, Kovarietal2019, Miszalskietal2019MNRAS487, Oroszetal2019}, 
for a small  sample of papers just from 2016 on), including also sub-stellar companions (e.g, \citealt{DeMarcoSoker2011, Boyle2018PhDT, SabachSoker2018}). These studies support the notion that single stars cannot account for even a small fraction of the rich variety of shapes of PNe (e.g., \citealt{SokerHarpaz1992, NordhausBlackman2006, GarciaSeguraetal2014}). 

The increasing recognition that binary interaction is behind the shaping of non-spherical PNe developed alongside the increasing recognition that jets play a crucial role in the shaping process, e.g., the companion star accretes mass and launches jets (e.g., \citealt{Morris1987, Soker1990AJ, SahaiTrauger1998, AkashiSoker2008, 
Boffinetal2012, HuarteEspinosaetal2012, Balicketal2013, Miszalskietal2013, 
Tocknelletal2014, Huangetal2016, Sahaietal2016, RechyGarciaetal2017, Derlopaetal2019, Tafoyaetal2019}, 
out of many more papers).
The interaction of jets with the envelope on the outer parts might lead to an optical transient (e.g., \citealt{SokerGilkis2018, YalinewichMatzner2019}). 
    
The recognition of the central roles of binary systems in stellar evolution brings with it the understanding that in many cases triple stellar systems might play a role (and not only in forming PNe, e.g., \citealt{MichaelyPerets2014, Portegies2016}).
Compared with the larger number of studies of binary interaction in forming and shaping PNe, only a small number of studies explore the presence of triple stellar systems in PNe and their possible role in shaping the PNe (e.g., \citealt{Sokeretal1992, Soker1994, Bondetal2002, Danehkaretal2013, Soker2016triple, BearSoker2017, Jones2017triple, Alleretal2018Triple, Gomezetal2018Triple, Jonesetal2019T, Miszalskietal2019T}). 
\cite{Exteretal2010} proposed that the AGB stellar progenitor of the PN SuWt~2 engulfed a tight binary system of two A-type stars, and that such an interaction  might form a nebula with a high-density equatorial ring (see also \citealt{Bondetal2002}). However, \cite{JonesBoffin2017} argue that the binary system of two A-type stars is unrelated to SuWt~2, and just happens to be along the line of sight to SuWt~2. 
Finally, we note that \cite{MichaelyPerets2019} and \cite{Igoshevetal2019} use triple stellar system to study post-common envelope evolution (CEE) systems and from that to infer the time scale of the CEE. 

In the present study we consider a tight binary system that enters the envelope of a giant star and experiences either a CEE or a grazing envelope evolution (GEE), or one after the other. 
We explore a new type of flow and base our simulation on two of our earlier studies, 
\cite{AkashiSoker2017} and \cite{Hilleletal2017PaperI}. 
Following a suggestion by \cite{Soker2004}, \cite{AkashiSoker2017} simulated the interaction of jets that a tight binary system launches into the wind of an AGB star. 
The orbital plane of the tight binary system is inclined to the orbital plane of the triple system, and so the jets that the tight binary system launches are not perpendicular to the orbital plane of the triple system. 
In general, when the orbital plane of the binary system is inclined to the orbital plane of the triple system the descendant PN will have no symmetry; neither point-symmetry, 
nor axial-symmetry, and nor mirror symmetry. Such PNe are termed `messy PNe' \citep{Soker2016triple, BearSoker2017}. 

In \cite{Hilleletal2017PaperI} we conducted three-dimensional (3D) hydrodynamical simulations where we assumed that the tight binary system merges inside the envelope of a red giant star and launches off-center jets for a short time. We deposited the energy of the merging process within a time period of nine hours, which is about one per cent of the the orbital period of the triple stellar system (the merger product around the center of the giant star). 
Such an interaction leads to the ejection of mass that when collides with previously ejected mass transfers kinetic energy to radiation, and therefore powers  an intermediate luminosity optical transient (ILOT).

Here we differ from the work of \cite{AkashiSoker2017}  by launching the two opposite inclined jets inside the giant envelope, and we differ from our earlier simulation by injecting inclined jets and for a long time period. As we discuss below, we might as well refer to the orbital motion of the tight binary system in the outskirts of the giant envelope as a GEE (e.g.,  \citealt{Shiberetal2017, Shiber2018}). 

We describe the 3D numerical code and the initial setting of the triple stellar system and jets in section \ref{sec:numerical}. In section \ref{sec:Jets-Envelope} we present the interaction of the jets with the envelope, and in section \ref{sec:outflow} we present the outflow properties. We summarise the main results in section \ref{sec:summary}.

\section{NUMERICAL SET-UP}
 \label{sec:numerical}
   
In the scenario that we simulate in the present study a tight binary system enters the envelope of an AGB star, namely, we have a CEE of a triple stellar system. The orbital plane of the tight binary system is inclined to the orbital plane of the triple stellar system, i.e., that of the tight binary system around the AGB star. We assume that one of the stars of the tight binary system accretes mass from the envelope and launches jets that are more or less perpendicular to the orbital plane of the tight binary system  \citep{Soker2004, Soker2016triple, AkashiSoker2017}. This means that the jets are inclined to the orbital plane of the triple system. In our simulations we take the inclination to be $45^\circ$ to emphasise the role of the tilted jets. The direction of the jets' axis is fixed relative to background stars. 
    
Many parts of the present simulation is similar to that in our earlier paper \cite{Hilleletal2017PaperI}. 
We use the stellar evolution code \texttt{MESA} \citep{Paxtonetal2013, Paxtonetal2015, Paxtonetal2018} to evolve a zero-age-main-sequence star of $M_{ZAMS}=4 M_\odot$ for $3\times10^8 \yr$, when it becomes an AGB star with a radius of $R_{AGB}=100\,R_{\odot}$ and an effective temperature of $T_{eff}= 3400K$. We take this AGB stellar model and install it at the center of our computational grid of the three-dimensional (3D) hydrodynamical code {\sc pluto} \citep{Mignone2007}. The center of the AGB does not change and it is at $(x_A,y_A,z_A)=(200,200,200)R_\odot$. 

To save expensive computational time we replace the inner $5\%$ in radius ($5\,R_{\odot}$) of the star with a sphere having a constant density, pressure, and temperature. The gravitational field in our simulation is constant in time and equals to that of the initial AGB star. Namely, we ignore both the gravity of the tight binary system and of the deformed envelope.

The computational grid is a cube with a side of $L_g=400 R_{\odot}$ with a base grid resolution of $\Delta L_g=L_g/48=5.8\times 10^{11} \cm$, and with an adaptive-mesh-refinement (AMR) of either 4 (highest resolution of $\Delta  L_g/2^4$; we refer to this as high resolution) or 3 (highest resolution of $\Delta L_g/2^3$; we refer to this as low resolution) refinement levels. In the entire computational grid the equation of state is that of an ideal gas with adiabatic index of $\gamma=5/3$.

We start the simulation at $t=0$ with the jets' source (the tight binary system) at $(x_b,y_b,z_b)_0=(x_A,y_A,z_A)+(50 R_\odot,0,0)=(250,200,200)R_\odot$, and let it otbit the center of the AGB on a circle with a constant radius of $a=50\, R_\odot$ and with a constant velocity equals to the initial local Keplerian velocity.
The AGB mass inner to that radius is $M(50 R_\odot)=2.7 M_\odot$, and the initial density at the orbit is $\rho(50 R_\odot)=1.3 \times 10^{-5} \g \cm^{-3}$. 
The orbital period is about 25 days, as that of a test particle at that radius. Including the mass of the tight binary system will give a somewhat shorter orbital period and a somewhat higher relative velocity between the tight binary system and the envelope. As we ignore the envelope rotation that reduces the relative velocity, we use the relative velocity of a test particle at that radius.  

For these parameters the Bondi-Hoyle-Lyttleton mass accretion rate of the tight binary system is $\dot M_{\rm 23,BHL} \approx 10 M_\odot \yr^{-1} [(M_2+M_3)/0.5M_\odot]^2$, 
where $M_2+M_3$ is the total initial mass of the tight binary system.
{{{{ In this study we do not simulate the tight binary system, and therefore the separate masses $M_2$ and $M_3$ are not relevant parameters. We do note that such low mass stars, $M_2 \simeq M_3 \simeq 0.2-0.3 M_\odot$, have deep convective envelope, something that ease the accretion of a relatively large amount of mass.  
}}}}

Although high mass accretion rates are possible inside a common envelope (e.g.,  \citealt{Chamandyetal2018a}), this is a too high mass accretion rate for low mass main sequence stars, and we assume that the actual mass accretion rate is only about one percent of that rate, namely $\dot M_{\rm 23,acc} \approx 0.1 [(M_2+M_3)/0.5M_\odot]^2 M_\odot \yr^{-1} $. We will take the mass outflow rate in the jets to be about ten per cent of the accretion rate.   
We start to launch the $45^{\circ}$ inclined jets also at $t=0$. 

We launch two opposite high-Mach number jets from their source in the following way. We take the initial velocity and kinetic energy of the two jets to be $v_j=500 \km \s^{-1}$, and $ \dot E_{\rm 2j}=10^{39} \erg \s^{-1}$, i.e., a mass flow rate into the two jets of $\dot M_{\rm 2j}=0.013 M_\odot \yr^{-1}$. 
We set two opposite cones at $45^\circ$ to the orbital plane of the simulation. Each cone has a length of $5 R_\odot$ and a half opening angle of $\alpha_j=30 ^\circ$. At each time step we add to each cell in these cones the mass and momentum ejected by the jets within the time step.

{{{{ The ram pressure of the jets for these parameters is $P_{\rm j,ram} = \rho_j v^2_j = 2.0 \times 10^8 (z_{\rm j}/ 5 R_\odot)^{-2} \erg \cm^{-3}$, where $z_{\rm j}$ is the distance of the jets' material from its origin (the tight binary system). The thermal pressure of the undisturbed envelope at the location of the tight binary system is $P_{\rm env} = 2.8 \times 10^8 \erg \cm^{-1}$. This implies that for $\alpha_j=30 ^\circ$ the ram pressure of the jets is lower than the envelope thermal pressure when $z_{\rm j} \ga 4 R_\odot$. In reality, it is reasonable to assume that at their origin the jets are narrower, say $\alpha_{\rm j} \simeq 15^\circ -20^\circ$, so that their ram pressure is larger than the envelope thermal pressure up to $z_{\rm j} =5 R_\odot$. At around $z_{\rm j} \simeq 5 R_\odot$, where envelope pressure becomes significant, the jets widen. Therefore, when the jets reach $z_{\rm j} \simeq 5 R_\odot$ their opening angle is $\alpha_{\rm j} \simeq 30^\circ$. These are the properties of the jets we inject in two opposite cones up to a distance of $z_{\rm j} =5 R_\odot$.   }}}}

As we discussed in our earlier paper \citep{Hilleletal2017PaperI}, for numerical reasons we ignore the deformation and the spin-up of the AGB envelope by the tight binary system before the tight binary system reaches the orbital separation $a=0.5 R_{\rm AGB} =50 R_\odot$.
The high-resolution run with 5 refinement levels was run for less than one orbital period because of a very long computational time. 
We degraded the resolution to 4 refinement levels to run several orbital periods. 
 
\section{Jet-envelop interaction and implications}
 \label{sec:Jets-Envelope}

We start our simulation ($t=0$) with the tight binary system already inside a spherically non-rotating envelope. This is not consistent with the expected evolution before that time where the companion entered the envelope and already spun-up and disturbed the envelope by its gravity and by launching jets. Therefore, we let the flow to build itself for half an orbital period, and only then present the flow structure. 
 
In the figures to come, the location of the center of the AGB star is at $(x_A,y_A,z_A)=(200,200,200)R_\odot$, and the initial location of the tight binary system at $t=0$, $(x_b,y_b,z_b)_0=(250, 200, 200)R_\odot$, is marked either with a black or with a white dot. The units on the axes of all figures are $R_\odot$. 

\begin{figure} 
\centering
{\includegraphics[width=0.40\textwidth]{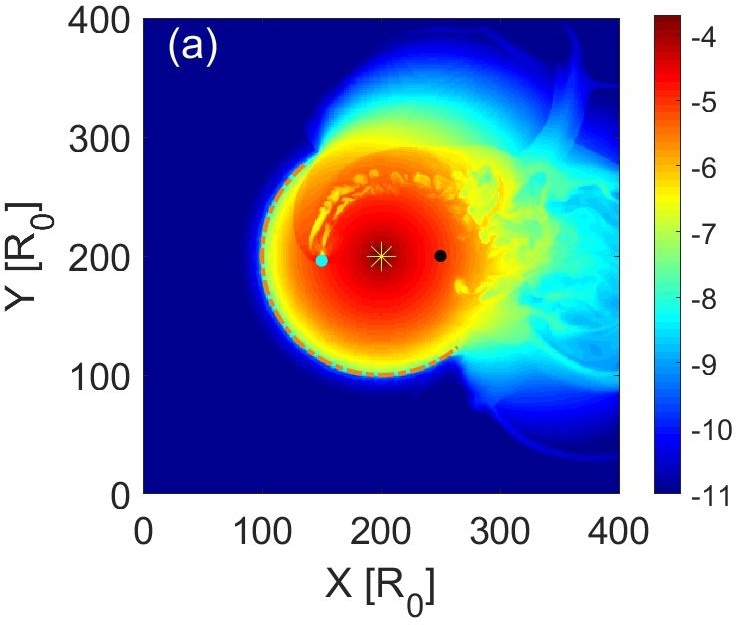}}
{\includegraphics[width=0.40\textwidth]{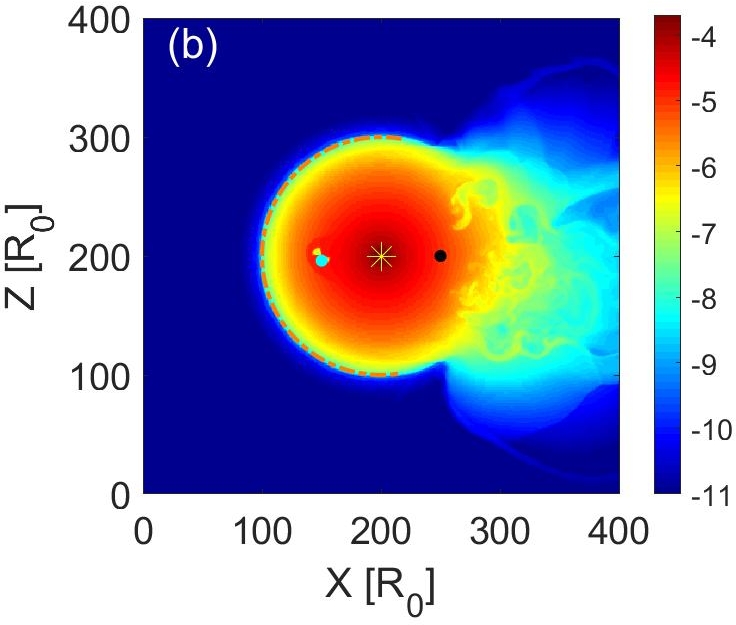}}
\caption{Density maps at $t=12.5 {~\rm days}$ of the high resolution simulation. The upper panel is in the orbital plane $z=0$, and the lower panel is in the meridional plane $y=0$. The left side of the lower panel shows the tilted
jets as the jets' source (the tight binary system) emerges from the $-y$ direction. Black dot is the initial location of the tight binary system that launches the jets, and the cyan dot is its present location. 
The density is according to the colour bar in log scale in $\g \cm^{-3}$.
} 
\label{fig:XY-Z-Density13}
\end{figure}

In Fig. \ref{fig:XY-Z-Density13} we present density maps at $t=12.5~$day, when the tight binary system has finished the first half of the first Keplerian orbit.
The upper panel shows the density in the equatorial plane $z=z_A=200R_\odot$ (which is also the orbital plane of the triple stellar system), and the lower panel shows it in the meridional plane $y=y_A=200R_\odot$. 
In the upper panel the tight binary system moves counterclockwise around the centre of the giant star, and we mark its location at any given time with a cyan dot. The dashed-dotted orange line marks the initial surface of the giant star $R_{\rm AGB}=100 R_\odot$.

\begin{figure} 
\centering
{\includegraphics[width=0.40\textwidth]{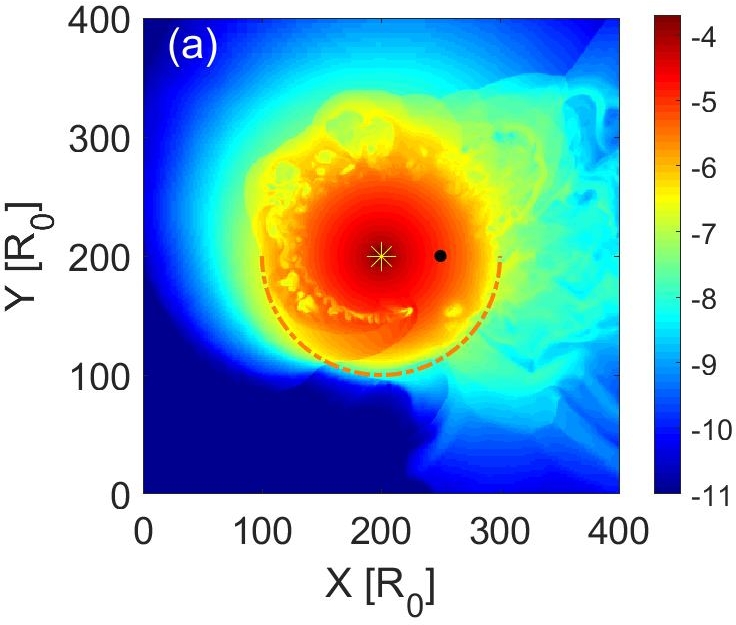}}
{\includegraphics[width=0.40\textwidth]{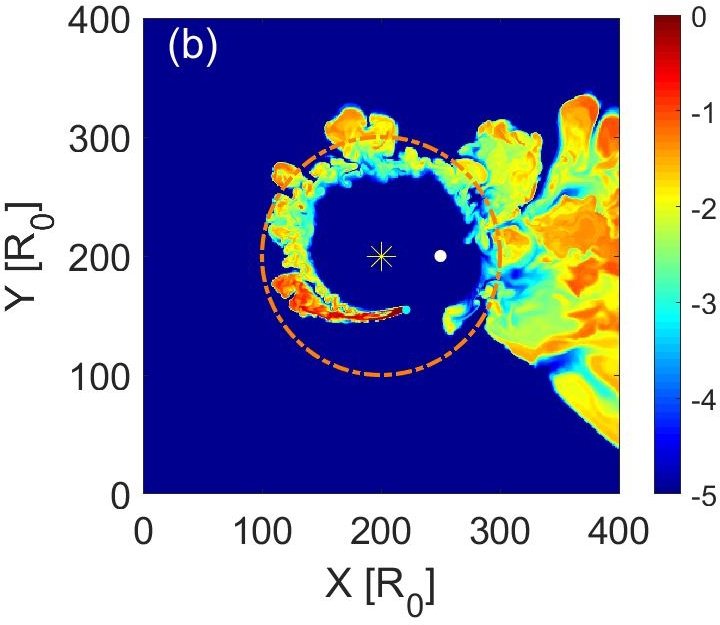}}
{\includegraphics[width=0.40\textwidth]{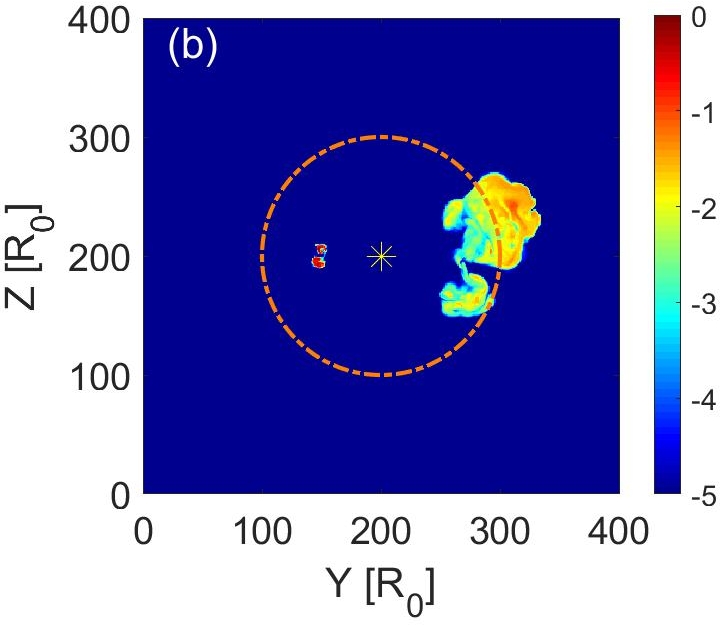}}
\caption{Maps of the density in the $z=0$ plane (upper panel; values are according to the colour bar in units of $\g \cm^{-3}$ and in log scale) and jet-tracer maps (values are according to the colour bars in log scale) at $t=20.4 {~\rm days}$ of the high resolution simulation. The jet-tracer marks the mass fraction of the gas that originated in the jets. 
The middle panel is the jet-tracer map in the $z=z_A+5 R_\odot=205R_\odot$ plane, and the lower panel is in the meridional plane $x=x_A=200R_\odot$. 
The two small red zones on the left of the lower panel shows the jets shortly after they were launched; the tight binary system that launched the jets has just crossed the $x=x_A$ plane in the $+x$ direction (just after three quarters of the first orbit). 
} 
\label{fig:XY-Z-Tracer14}
\end{figure}

To follow the gas that we inject into the jets we use a `jet-tracer', which marks the fractional mass of gas that comes from the jets in each numerical cell. Where we inject the jets the value of the jet-tracer is 1, and in numerical cells where there is no gas that comes from the jets the value of the jet-tracer is zero. In Fig.  \ref{fig:XY-Z-Tracer14}  we present the jet-tracer maps in two planes, as well as the density, at $t=20.4$~day.

From Figs. \ref{fig:XY-Z-Density13} and \ref{fig:XY-Z-Tracer14} we learn the following. We see a large outflowing `cloud' on the positive x-axis (right side). But, as we stated above, as our flow at early times is not consistent with earlier expected evolution, we do not consider this outflow here. 
The jets material at later times does not spread much (Fig. \ref{fig:XY-Z-Tracer14}), but the pressure that the shocked jets' gas builds accelerates the envelope gas out in all directions. We will return to this flow pattern in section \ref{sec:outflow}.  
The tracer map in the $z=z_A+5 R_\odot$ plane shows a chain of vortexes. These trailing vortexes are formed by a process called vortex shading, and they form the von Karman vortex street behind the jets. The density maps show these vortexes to have low densities. 
Because jet-induced vortexes can have an important role both in the dynamics of the flow and in heat transport (e.g., \citealt{HillelSoker2016}), 
we concentrate on these.
    
\begin{figure} 
\centering
\includegraphics[width=0.50\textwidth]{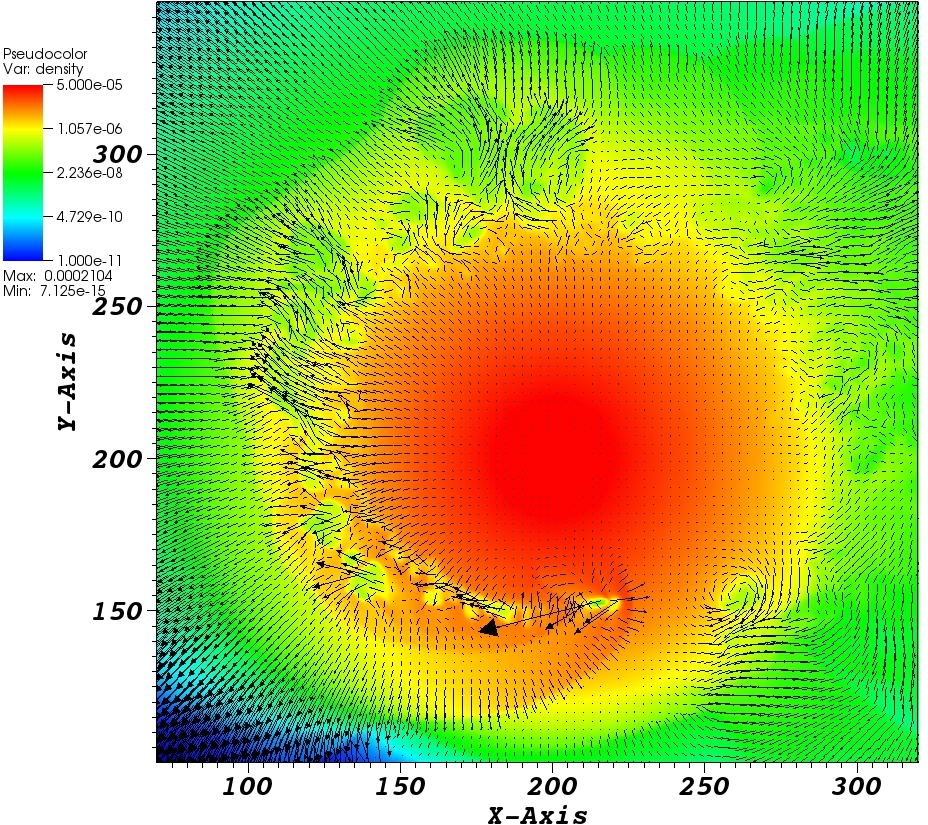}
\caption{Density map and arrows that depict the flow velocity, magnitude and direction, in the equatorial plane at $t=20.4 {~\rm days}$ of the high-resolution simulation. 
The density map is as in the upper panel of Fig. \ref{fig:XY-Z-Tracer14} but with a different colour scale and a smaller region. Density according to the colour bar from $10^{-11} \g \cm^{-3}$ (blue) to $5 \times 10^{-5} \g \cm^{-3}$ (red). The length of each arrow is proportional to the velocity it represents, with a maximum velocity of $300 \km \s^{-1}$. 
Note the vortexes that the jets shed along the orbit.  
} . 
\label{fig:Vortexes1}
\end{figure}

In Fig. \ref{fig:Vortexes1} we present the density map as in the upper panel of Fig. \ref{fig:XY-Z-Tracer14} (but with a different colour scale and a smaller portion of the equatorial plane), including now arrows that depict the flow pattern in the equatorial plane. The velocity arrows clearly show the many vortexes that the jets shed. These vortexes trail behind the tight binary system (the source of the jets) as it orbits inside the AGB envelope. 
   
\begin{figure} 
\centering
\includegraphics[width=0.447\linewidth]{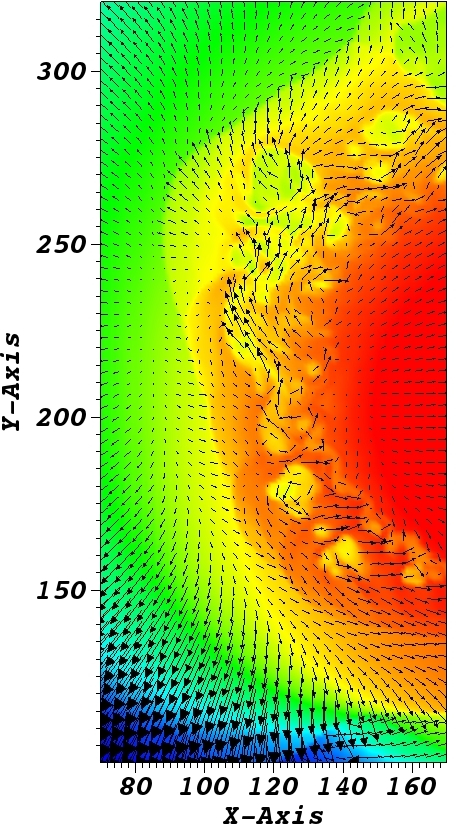}
\includegraphics[width=0.41\linewidth]{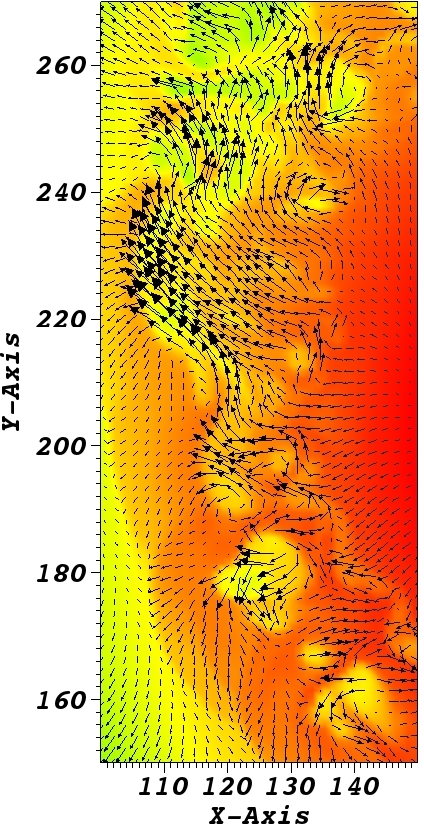}
\caption{Similar to Fig. \ref{fig:Vortexes1} (high resolution run at $t=20.4 {~\rm days}$; same colour coding), but focusing on smaller parts of the left side of the equatorial plane ($x<0$). Densities from $10^{-11} \g \cm^{-3}$ (blue) to $10^{-5} \g \cm^{-3}$ (red). The maximum velocity in the left and right panels are $320 \km \s^{-1}$ and $300 \km \s^{-1}$, respectively. 
To emphasise the vortexes in the flow expanding to the left we added a velocity of $v_x=30.8 \km \s^{-1}$ and $v_x=15.4 \km \s^{-1}$ in the left and right panels, respectively. Note the different length scales of the two panels.  
} 
\label{fig:Vortexes2}
\end{figure}

In Fig. \ref{fig:Vortexes2} we focus on the left hand side of Fig. \ref{fig:Vortexes1} ($x<x_A$ side of the orbital plane). In these regions the expanding flow has a negative x-component velocity, $v_x<0$. Therefore, to emphasise the vortexes in the flow that expands to the left we added a positive x-component velocity to each of the panels. In the left panel we added a velocity of $v_x=30.8 \km \s^{-1}$ and in the right panels a velocity of $v_x=15.4 \km \s^{-1}$. 

\begin{figure} 
\centering
{\includegraphics[width=0.45\textwidth]{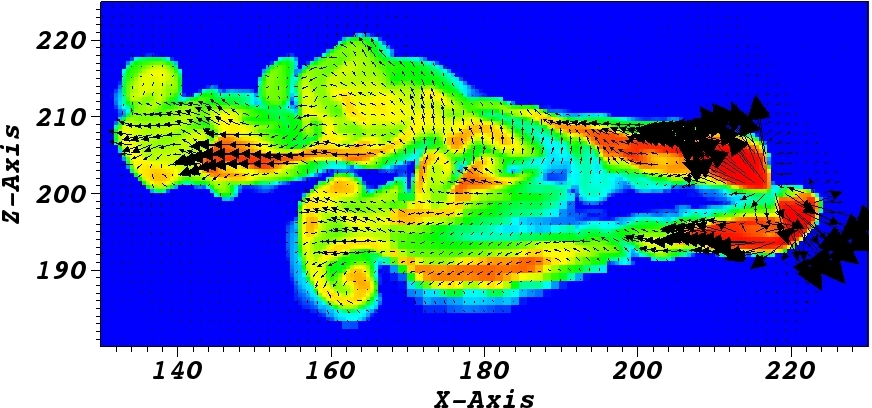}}
{\includegraphics[width=0.45\textwidth]{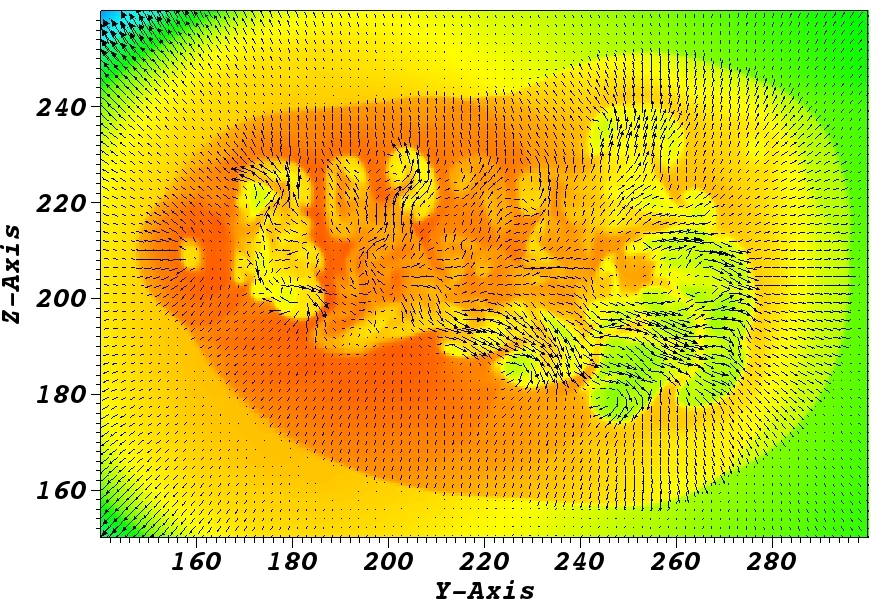}}
\caption{Focusing on small regions along the orbit of the tight binary system in the high resolution run at $t=20.4 {~\rm days}$. 
Upper panel: The jet-tracer and velocity maps in the $y=153R_\odot$ plane containing the jets' source and the region behind it. Colour coding is as in the lower panels of Fig. \ref{fig:XY-Z-Tracer14}. The large arrows are the tilted jets. The tight binary system is between the two red regions (pure jets' material)  at $(x_b,y_b,z_b)=(216, 153,200)R_\odot$. 
Lower panel: The density and velocity in the $x=125R_\odot$ plane that covers about the same region as in Fig. \ref{fig:Vortexes2}, but in a perpendicular plane. Densities from $10^{-11} \g \cm^{-3}$ (blue) to $10^{-5} \g \cm^{-3}$ (red), with the same colour coding as in Fig. \ref{fig:Vortexes1}. The maximum velocity in the upper and lower panels are $480 \km \s^{-1}$ and $295 \km \s^{-1}$, respectively. Note the different length scales of the two panels.
} 
\label{fig:Vortexes3}
\end{figure}

When a jet propagates along a constant axis the vortexes trail behind its head and along its axis. Here the axis of the jets moves around the center of the AGB star. As we mentioned above, the jets induce vortexes that trail behind and along the orbit of the jets' source (the tight binary system). To show the vortexes just behind the source of the jets, in the upper panel of Fig. \ref{fig:Vortexes3} we focus on the flow behind the tight binary system in the $y=153 R_\odot$ plane. We present the jet-tracer map and a velocity map by arrows. The length of each arrow is proportional to the velocity it represents. The tight binary system is between the two red regions at $(x_b,y_b,z_b)=(216, 153,200)R_\odot$. 
In the lower panel we present the vortexes along the orbit but at about a quarter of an orbit behind the tight binary system. We present the density and velocity maps in the $x=125R_\odot$ plane, covering about the same region as in Fig. \ref{fig:Vortexes2}.
 Both panels are at $t=20.4 {~\rm days}$. 

{{{{ The typical diameter of the vortexes is $\approx 10 R_\odot$. This is about the size of the region from where we inject the jets and twice the cross section of the jets at their injection radius $2r_{\rm j} = 2 z_{\rm j} \sin \alpha_j=5 R_\odot$. Our experience with the hydrodynamical code {\sc pluto} \citep{RefaelovichSoker2012} shows that the vortexes are somewhat larger than the cross section of the jets (but not too small as there is a numerical viscosity that suppresses small vortexes). We also found there that higher velocity jets induce somewhat smaller vortexes (but the dependence on velocity is weak). }}}}

{{{{ The envelope of cool giant stars is convective. Near the origin of the jets in our simulation the ram pressure of the jets, and therefore the thermal pressure of the post-shock jets' material, is larger than the ram pressure of the convective motion (because the convection is subsonic). We expect therefore that adding a convection to the envelope in the simulation will not affect much the formation and early evolution of the vortexes. As the spiral structure behind the orbiting jets' origin (the tight binary system here) widens, the convective motion will influence more the evolution of the vortexes, up to erasing the clear structure of the vortexes.  
}}}} 

All figures of this section show that the density fluctuations accompany the vortexes. The turbulent velocity pattern and the density fluctuations of the vortexes might have three effects on the outflow. 
First, the vortexes that the jets form make energy transport much more efficient than that of the convection of a single AGB star. This energy transport can efficiently carry the recombination energy of hydrogen out from the envelope so it is radiated away rather than contributing to envelope removal (e.g.,  \citealt{Sabachetal2017, Grichener2018, WilsonNordhaus2019}). 
     
The second effect is related to the first as it involves heat transport. \cite{Chamandyetal2019} argue that the heat transport by mixing distributes orbital energy that the in-spiral binary or triple system releases, and by that it increases the efficiency of common envelope removal (also \citealt{WilsonNordhaus2019}). 
\cite{Shiberetal2019} find hints that the jets in their 3D simulations of the CEE indeed redistribute the orbital energy and the jets' energy, making envelope removal more efficient. The above mentioned new studies motivate us to present detailed velocity maps, which show indeed the spreading of vortexes in the envelope. The vortexes distribute in the common envelope not only the orbital energy, but also the energy that the jets themselves deposit \citep{Shiberetal2019}.

The third possible effect of the vortexes might result from the density fluctuations that accompany the vortexes. It is possible that after mass flows out to cooler regions the  denser zones facilitate early dust formation. As we do not include here neither radiative transfer nor radiation pressure on dust, we leave this effect to future studies, as dust seems to be an important ingredient in the last phases of the CEE (e.g., \citealt{Soker1992, GlanzPerets2018, Iaconietal2019}).

In a recent study \cite{Iaconietal2019} find that as the ejecta of a CEE expand away from the binary system the turbulent distribution in both density and velocity slowly becomes more globally symmetric. So late observations cannot reveal the fine details of the vortexes and clumps as we present here.

The main result of this section is the demonstration, through detailed velocity and density maps, of the complicated structure of vortexes that the jets induce in the common envelope. These vortexes might have an important role in envelope ejection. 

\section{Outflow properties}
 \label{sec:outflow}
 
In this section we describe some properties of the large scale outflow. First we list once more a few of the limitations of our simulations. 
(1) We start with the tight binary system already inside the envelope. For that we do not concentrate at early times, i.e., the first half of an orbital period (although we do not ignore that period).  
(2) We do not calculate the accretion rate at each time, and so we cannot estimate the evolution of the jets' power. We rather use a constant jets' power as we explained in section \ref{sec:numerical}. 
(3) We do not consider the spiralling-in process, but rather inject the jets at an orbit with a constant radius $a=50 R_\odot$. To calculate self-consistently the orbit we need to include self-gravity of the envelope and the gravity of the tight binary system, which would make jets' inclusion very computer-demanding (one of us, Noam Soker, applied for a grant to conduct such a study over 4 years, but was turned down few months ago). We partially justify the constant orbit by the following two considerations.  

From recent 3D hydrodynamical simulations of the CEE we expect the tight binary system to rapidly spiral-in to $\approx 5-20 R_\odot$ (e.g., \citealt{IvanovaNandez2016, Ohlmannetal2016, Iaconietal2017, Reichardtetal2019, Shiberetal2019}). We therefore take the average orbit to be $\approx 50 R_\odot$. The second consideration is that mass removal by the jets slows down the spiralling-in process, and may even stop it, e.g., the GEE. 
Overall, we isolate the influence of the tilted jets on the outflow, i.e., not including the self gravity of the envelope and the orbital energy on the mass ejection. 

{{{{ We further note that we do not ignore any gravitational energy of the giant star. We only assume that its gravitational potential does not change with time. This assumption might actually reduce the envelope unbinding efficiency. In reality, when the envelope expands as a result of energy deposition the gravity at each radius inside the envelope decreases. Therefore, in the next round the mass that the jets interact with sits in a shallower potential well, easing mass removal compared with the case we simulate here. }}}} 

\begin{figure} 
\centering
{\includegraphics[width=0.40\textwidth]{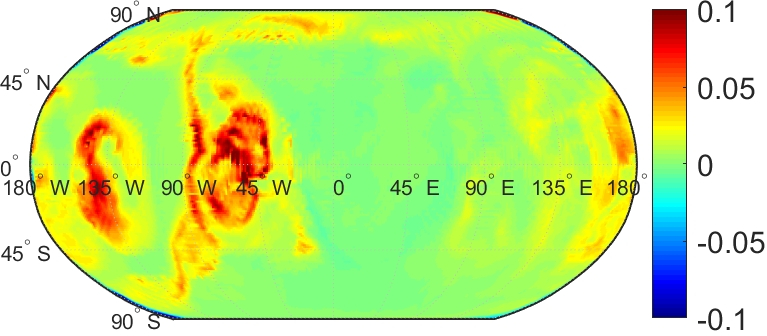}}
{\includegraphics[width=0.40\textwidth]{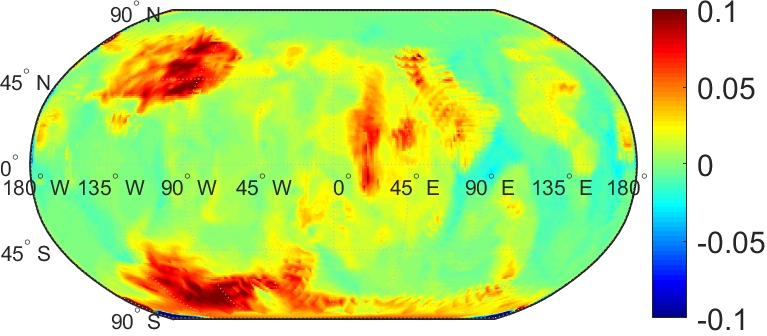}}
{\includegraphics[width=0.40\textwidth]{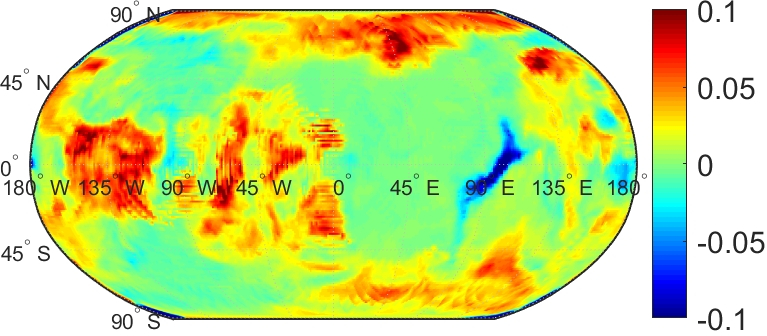}}
{\includegraphics[width=0.40\textwidth]{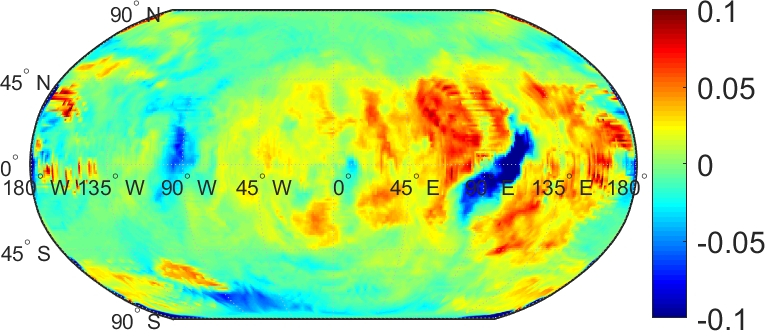}}
\caption{Mass flux through a sphere with a radius of $R_{\rm out}=190 R_{\odot}$ in the low-resolution simulation. Mass flux is in units
of $\g \cm^{-2} \s^{-1}$ as indicated by the colour bars. The simulation times are (top to bottom) 38, 51, 64 and 77 days, which amounts to about a half an orbital period between two consecutive panels. Negative values imply an inflow.
The initial position of the tight binary system is at direction $(0,0)$. 
} 
\label{fig:MassLoss}
\end{figure}

We first examine the momentarily mass flux in each radial direction, $\phi=\rho v_r$, that crosses a sphere of radius $R_{\rm out}=190 R_\odot$ centred on the center of the AGB star (the center of the grid). 
In Fig. \ref{fig:MassLoss} we present the mass flux at four times, separated by about half an orbital period, using a projection of the entire sphere. We note that in some areas there is an inflow (blue colour).

\begin{figure} 
\centering
{\includegraphics[width=0.40\textwidth]{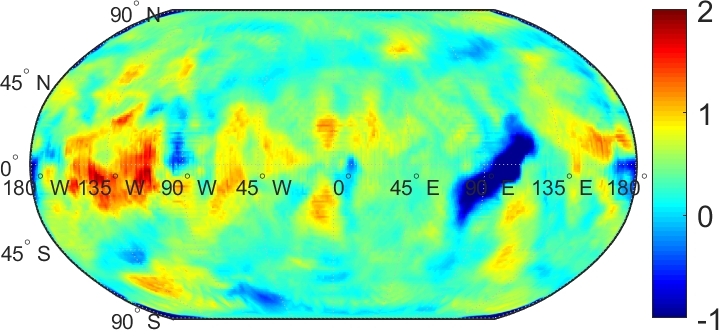}}
\caption{Colour map of the quantity $\Phi=4 \pi R^2_{\rm out} \overline \phi$ in units of $M_\odot \yr^{-1}$. It is the average mass loss rate as if the entire sphere would have the same average mass flux $\overline \phi$ during the time period $35-85~{\rm days}$. The initial position of the tight binary system is at direction $(0,0)$.}
\label{fig:MassFlux}
\end{figure}

In Fig. \ref{fig:MassFlux} we present the average mass flux from this sphere during a time period of two orbits, from 1.4 orbits ($t=35~{\rm days}$) to 3.4 orbits ($t=85~{\rm days}$). We actually present the quantity $\Phi=4 \pi R^2_{\rm out} \overline \phi$ that has units of $M_\odot \yr^{-1}$. It is the average mass loss rate as if the entire sphere would have the same average mass flux $\overline \phi$. 
During that time period the total mass that was lost is $\Delta M_{\rm out}=0.0037 M_\odot$, which gives an average mass loss rate of $\dot M_{\rm jets}=0.027 M_\odot \yr^{-1}$. This mass loss rate is due solely to the effect of the jets. The total mass that was injected into the jets during the time period from $t=0$ to $t=85~{\rm days}$ is $\Delta M_{2j}=0.003 M_\odot$, but most of it is still in the grid. 

Due to the limited scope of our simulations (as we explained above), we cannot yet reach broad conclusions. However, two conclusions from the results we present in this section seem robust. (1) As earlier numerical simulations of CEE or GEE with jets showed, both those that included self gravity of the distorted envelope \Citep{Shiberetal2019} and those that did not include (e.g., \citealt{Shiber2018}), jets are efficient in removing mass from the common envelope.
That we achieve this mass removal with an assumed accretion rate that is only about one per cent of the Bondi-Hoyle-Lyttleton mass accretion rate, strengthens the claim that jets might play an important role in mass removal during the CEE and the GEE. 
(2) The outflow is very clumpy. As we discussed in section \ref{sec:Jets-Envelope}, such clumps might facilitated the formation of dust, that with the radiation of the system might play an important role in the mass loss process of the CEE. Of course, other AGB stars can also form dusty clumps, such as when there is a binary companion outside the envelope (e.g., \citealt{Decinetal2015}). 

\section{SUMMARY}
 \label{sec:summary}

We studied a rare type of evolutionary route where a tight binary system, most likely composed of two low-mass main sequence stars, enters the envelope of a giant star. We assumed that the tight binary system accretes mass and at least one of the stars launches jets. We further assumed that the orbital plane of the tight binary systems is inclined to the orbital system of the tight binary system with the giant star, such that the jets are inclined to the orbital plane of the triple system. 

Although this type of interaction is very rare, it adds to other rare types of interactions that might teach us on more abundant types of triple star interactions and on the CEE, and might explain some rare messy morphologies of PNe.
Our simulations are complementary to the study of \cite{AkashiSoker2017} who simulated the interaction of inclined jets with the wind of the giant star when the tight binary system that launches the jets is outside the envelope, and to our earlier study of the merger of the tight binary system inside the envelope \citep{Hilleletal2017PaperI}. Like the earlier two studies, we concentrate on the role of the jets. 
The triple stellar interaction of the three studies might form messy PNe, i.e., PNe lacking any type of symmetry (see section \ref{sec:intro}). 

In this first study of such a flow we assumed a constant orbital separation of the tight binary system inside the envelope, we ignored the spin-up and deformation of the giant envelope before the starting of our simulations, and we ignored the self gravity of the envelope and the gravity of the tight binary system. Overall, we isolated the roles of the jets.

We described our main results of the small scale interaction in section \ref{sec:Jets-Envelope} and of the large scale flow in section \ref{sec:outflow}. 
The jets induce many vortexes in the common envelope (Figs. \ref{fig:XY-Z-Density13}-\ref{fig:Vortexes3}).
As we discussed in section \ref{sec:Jets-Envelope}, these vortexes can efficiently transport the recombination energy, which is thermal energy, to be radiated away on the one hand, but redistribute the orbital energy and the energy carried by the jets in the envelope, both mechanical energies, and by that to support envelope removal on the other hand
(e.g., \citealt{Chamandyetal2019, WilsonNordhaus2019, Shiberetal2019}). 
The dense zones of the clumpy outflow that result from the vortexes might facilitated dust formation in the outflowing envelope. The role of dust in the CEE deserve more detail studies. 

In short, the large scale outflow is `messy'. However, unlike the case of jets that a tight binary companion launches outside the envelope \citep{AkashiSoker2017} and that inflate large bubbles in the wind of the giant star, here the jets do not inflate large bubbles. The outflow morphology we obtained here might explain PNe having messy morphologies without large bubbles and arcs. Examples (see \citealt{BearSoker2017}) might be the PN H~2-1 (PN~G350.9+04.4; e.g., image by \citealt{Sahaietal2011}), and, in a case of a very low mass tight binary system with a small influence on the morphology, NGC~7094 (PN G066.7−28.2; e.g., image by \citealt{Manchadoetal1996}).    

On a broader scope, our study adds to the rich variety of shapes that jets can form in nebulae around evolved stars, PNe and massive stars. The messy outflow we find here can be relevant to PNe as well as to rare luminous blue variables (LBVs). A recent demonstration of a morphological feature that is common to a PN and to at least one LBV and that jets might explain is a `column crown'. A column crown is a structure of many thin filaments protruding outward from the lobes of a bipolar structure  (the filaments might be along or inclined to radial directions). \cite{AkashiSoker2018} demonstrated that jets can form a columns crown. Both the PN Mz~3 (e.g., image by \citealt{Clyneetal2015}) and the LBV Eta Carinae \citep{SmithMorse2019} have a bipolar morphology with a columns crown on each of the two lobes. This strengthens the claim that jets shape the bipolar structure (the Homunculus) of Eta Carinae.    
Future observations might reveal nebulae around LBVs and other massive stars that have been shaped by jets in triple stellar systems.

\section*{Acknowledgments}
We acknowledge the great help of Efrat Sabach in building the AGB model.
{{{{ We thank an anonymous referee for very useful comments. }}}}
This research was supported by a generous grant from Prof. Amnon Pazy Research Foundation.
N.S. research is partially supported by the Charles Wolfson Academic Chair.


\label{lastpage}
\end{document}